\documentclass[twoside]{LCWS11}
\usepackage[latin1]{inputenc}
\usepackage[dvips]{graphicx,epsfig,color}
\usepackage{wrapfig,rotating}
\usepackage{amssymb,amsmath,array}
\usepackage{cite}
\begin{document}
\title{
Nonlinear Energy Collimation System for Linear Colliders} 
\author{Javier Resta-L\'opez
\thanks{resta@ific.uv.es}
\vspace{.3cm}\\
Instituto de F\'isica Corpuscular (IFIC), centro mixto CSIC-Universidad de Valencia\\
Institutos de Investigaci\'on de Paterna, Aptdo. 2285, 46071, Valencia, Spain 
}

\maketitle

\begin{abstract}
The post-linac energy collimation system of multi-TeV linear colliders is designed to fulfil an important function of protection of the Beam Delivery System (BDS) against miss-steered beams likely generated by failure modes in the main linac. For the case of the Compact Linear Collider (CLIC), the energy collimators are required to withstand the impact of a full bunch train in case of failure. This is a very challenging task, assuming the nominal CLIC beam parameters at 1.5~TeV beam energy. The increase of the transverse spot size at the collimators using nonlinear magnets is a potential solution to guarantee the survival of the collimators. In this paper we present an alternative nonlinear optics based on a skew sextupole pair for energy collimation. Performance simulation results are also presented.      
\end{abstract}

\section{Introduction}

The main functions of the post-linac collimation system for future linear colliders are: the cleaning of the beam halo particles which potentially can generate detector background at the Interaction Point (IP); and the protection of the Beam Delivery System (BDS) by minimising the activation and damage of sensitive accelerator components. 

In the BDS there are usually two collimation sections: one for energy collimation and other one for transverse phase collimation or betatron collimation. Here we focus our study on the energy collimation system. 

The energy collimation system is dedicated to collimate beam particles with large energy deviation. In addition, it can fulfil a very important protection function intercepting miss-steered or errant beams with energy offset generated in the main linac. This protection function is crucial for multi-TeV colliders, such as the Compact Linear Collider (CLIC), where energy errors generated by failure modes in the main linac are expected to be much more frequent than large betatron oscillations with small emittance beams.

The conventional collimation schemes are based on mechanical collimation using spoilers (scrapers) and absorbers. It could include several stages. For CLIC the energy collimation system includes a single spoiler-absorber scheme located in a region with horizontal dispersion. The nominal CLIC beam parameters and a complete description of the CLIC baseline linear collimation system can be found in \cite{Resta1}.

For CLIC the self-protection of the energy collimators is desirable, i.e. the energy collimators (spoiler and absorber) are required to withstand the impact of a full bunch train. This condition makes the energy collimator design very challenging. In order to guarantee the collimator survival, the following issues are currently being investigated: the study of novel materials with suitable electrical and thermo-mechanical properties, and the design of alternative optical layouts.   

In this paper we present an alternative optics design including nonlinear magnets to increase the spot size at the collimator position. Requirements and conditions for self-cancellation of optical aberrations are explained in Section~2. An optical layout of a nonlinear collimation system for CLIC is also presented in Section~2. The optical optimisation of this system is described in Section~3. Finally, performance simulation results are presented and discussed in Section~4.   

\section{Nonlinear passive protection}

\begin{wrapfigure}{r}{0.5\columnwidth}
\centerline{\includegraphics[width=0.45\columnwidth]{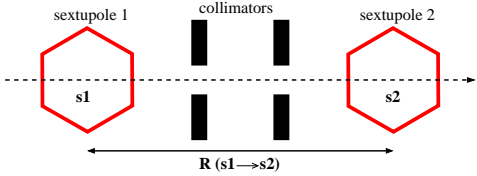}}
\caption{Basic schematic of a nonlinear collimation system based on a pair of skew sextupoles.}\label{basicscheme}
\end{wrapfigure}

By nonlinear passive protection we mean the use of a nonlinear magnet of moderate strength to increase the beam spot size at downstream mechanical collimators. Somehow this nonlinear element would play the role of a primary spoiler. A second nonlinear magnet of the same family is located downstream of the collimators to cancel the optical aberrations introduced by the former nonlinear element. Figure~\ref{basicscheme} shows a basic scheme $s1$-$R(s1\rightarrow s2)$-$s2$ of this nonlinear collimation concept based on two skew sextupoles $s1$ and $s2$.


\subsection{Optics design}

Using the transport formalism, the general optical constraints to cancel geometric nonlinear terms between two skew sextupoles are the following:

\begin{equation}
R_{12}=0,\,\,R_{34}=0,\,\,|R_{11}|=|R_{33}|, \,\,|R_{22}|=|R_{44}|\,\,,
\label{opticseq1}
\end{equation} 

\noindent where $R_{11}$, $R_{22}$, $R_{33}$, $R_{44}$, $R_{12}$ and $R_{34}$ are elements of the transverse first order transport matrix between the two sextupoles. In terms of the Twiss parameters $\beta_{x,y}$ and $\alpha_{x,y}$ at the sextupoles $s1$ and $s2$, and the phase advance $\mu_{x,y}(s1 \rightarrow s2)$,

\begin{eqnarray}
R_{11} & = & \sqrt{\beta_{x2}/\beta_{x1}}\left(\cos\mu_x(s1\rightarrow s2) + \alpha_{x1} \sin \mu_x(s1\rightarrow s2)\right)\,\,,\nonumber \\
R_{12} & = & \sqrt{\beta_{x1}\beta_{x2}}\sin \mu_x(s1\rightarrow s2) \,\,,\nonumber \\
R_{33} & = & \sqrt{\beta_{y2}/\beta_{y1}}\left(\cos\mu_y(s1\rightarrow s2) + \alpha_{y1} \sin \mu_y(s1\rightarrow s2)\right)\,\,,\nonumber \\
R_{34} & = & \sqrt{\beta_{y1}\beta_{y2}}\sin \mu_y(s1\rightarrow s2) \,\,,\nonumber \\
R_{22} & = & \sqrt{\beta_{x1}/\beta_{x2}}\left(\cos\mu_x(s1\rightarrow s2) - \alpha_{x2}\sin\mu_x (s1\rightarrow s2)\right)\,\,,\nonumber \\
R_{44} & = & \sqrt{\beta_{y1}/\beta_{y2}}\left(\cos\mu_y(s1\rightarrow s2) - \alpha_{y2}\sin\mu_y (s1\rightarrow s2)\right)\,\,.
\label{opticseq2}
\end{eqnarray}

\noindent From Eqs.~(\ref{opticseq1}) and (\ref{opticseq2}) one obtains $\mu_x(s1 \rightarrow s2)=n_x \pi,\,\,\mu_y(s1 \rightarrow s2)=n_y \pi$, where $n_x$ and $n_y$ are integers, and $\beta_{x2}/\beta_{x1}=\beta_{y2}/\beta_{y1}$.

The normalised integrated strength of the first skew sextupole, $K_{s1}$, is selected to get enough transverse beam spot size at the spoiler position for spoiler survivability in case of direct beam impact. For cancellation of geometric aberrations the second skew sextupole strength, $K_{s2}$, must satisfy

\begin{equation} 
K_{s2}=-K_{s1}R_{22}/R^2_{11} \,\,,
\end{equation}

\noindent which can be written in terms of betatron functions:

\begin{equation}
K_{s2}=(-1)^{1+n_y}K_{s1} \left(\beta_{x1}/\beta_{x2}\right)^{3/2}\,\,.
\label{opticseq3}
\end{equation}

By simplicity we use the $-I$ transfer matrix in both $x$ and $y$ planes between the sextupoles, which is a special case of the previous conditions. In this case, $n_x$ and $n_y$ are odd integers (by simplicity we select $n_x=n_y=1$), and $\beta_{x1}=\beta_{x2}$, $\beta_{y1}=\beta_{y2}$, $\alpha_{x1}=\alpha_{x2}$, $\alpha_{y1}=\alpha_{y2}$, $\mu_{x,y}(s1 \rightarrow s2)=\pi$. From these conditions and Eq.~(\ref{opticseq3}) one obtains $K_{s1}=K_{s2}$.

Additional optical constraints for the design of the nonlinear collimation system are the following:

\begin{itemize}
\item Non-zero dispersion at the collimator position. At the same time, the strength and length of the dipoles should be selected to avoid intolerable emittance growth due to incoherent synchrotron radiation effects (see Section~\ref{SReffects}). 

\item To cancel chromatic and chromo-geometric aberrations between the sextupole pair: $R_{16}(s1 \rightarrow s2)=0$, i.e. $D_{x1}=-D_{x2}$, with $D_{x1}$ and $D_{x2}$ the first order horizontal dispersion at the first and second skew sextupole respectively. 

\item At the end of the energy collimation section $D_x=0$ and $D'_x=dD_x/ds=0$, and we perform the matching with the betatron collimation section. 
\end{itemize}

\begin{wraptable}{r}{0.5\columnwidth}
\centerline{\begin{tabular}{|l|c|}
\hline
Parameter & Value \\
\hline \hline
Sext. strength $K_s$ [m$^{-2}$] & 8 \\
Product of pole-tip field & {} \\
and length $B_T \cdot l_s$ [T$\cdot$m] & 2 \\
Pole-tip radius $a_s$ [mm] & 10 \\
Effective length $l_s$ [cm] & 100 \\
Hor. beta function $\beta_{x,s}$ [m] & 436.6 \\
Vert. beta function $\beta_{y,s}$ [m] & 110.2 \\
Hor. dispersion $|D_{x,s}|$ [m] & 0.097 \\
\hline
\end{tabular}}
\caption{Skew sextupole parameters.}
\label{skewsextparam}
\end{wraptable}

\begin{figure}
\centerline{\includegraphics[width=0.5\columnwidth]{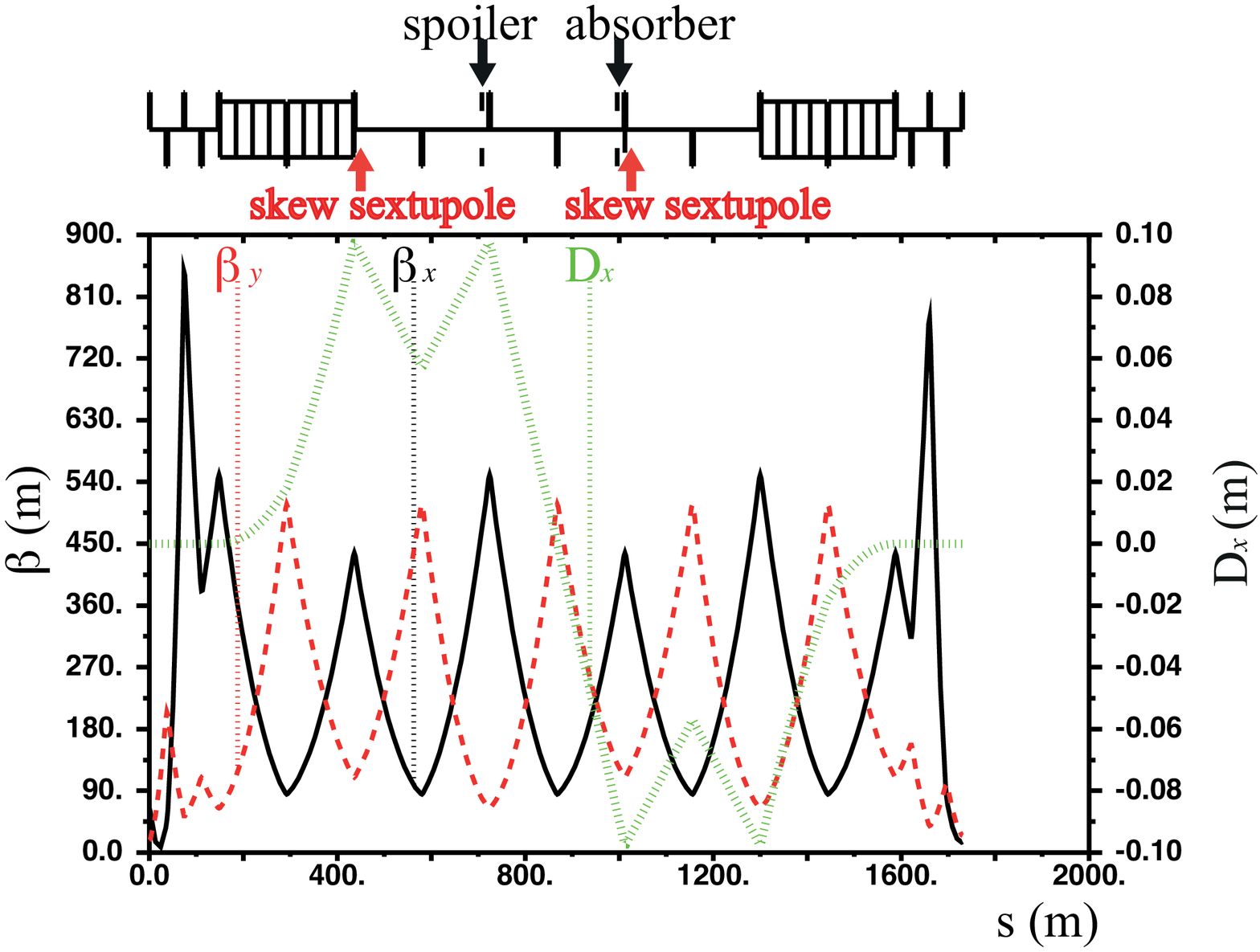}
\includegraphics[width=0.5\columnwidth]{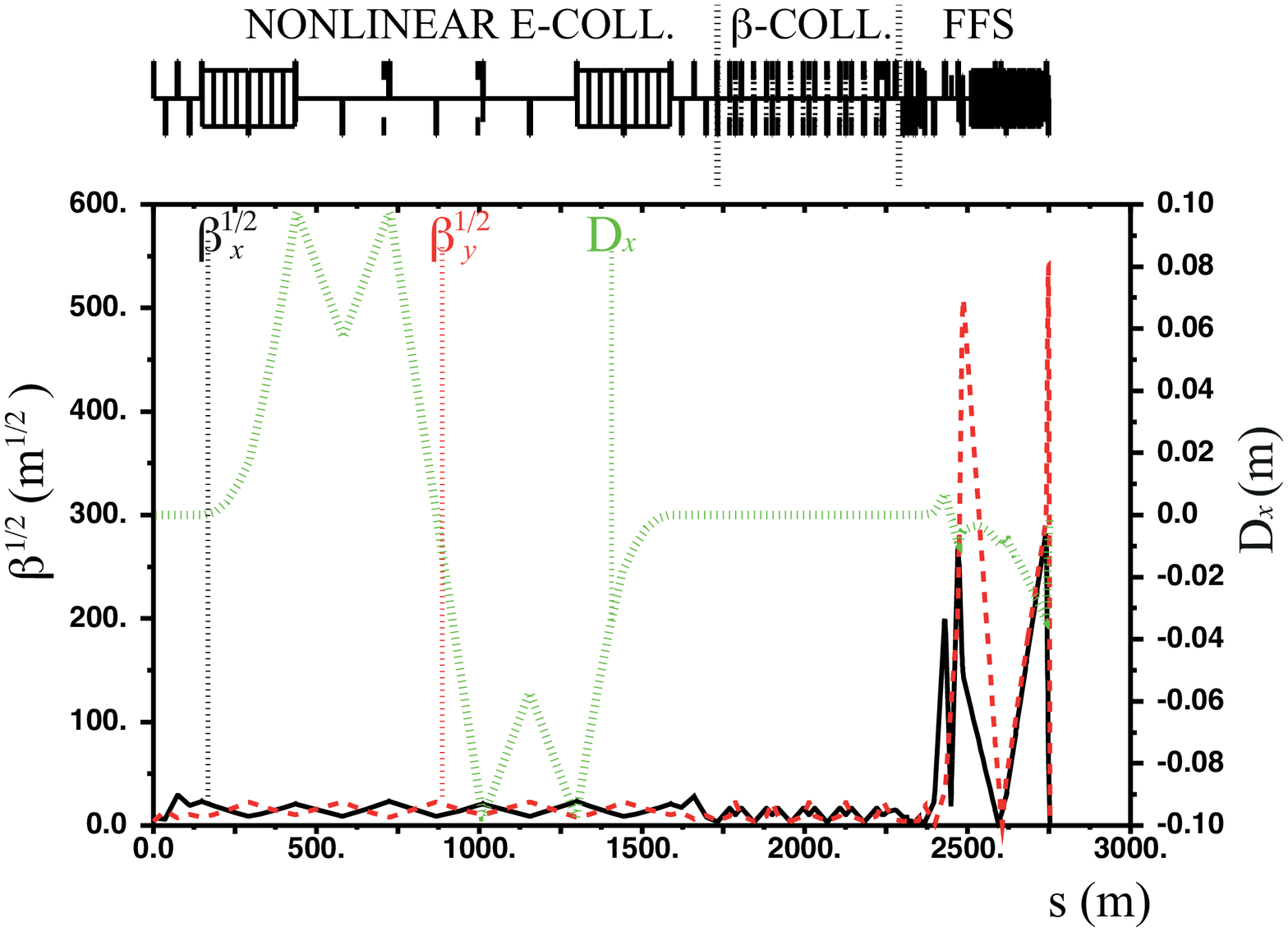}}
\caption{Left: layout and optical functions of a nonlinear energy collimation system for CLIC. Right: layout and optical functions of the CLIC BDS including the nonlinear energy collimation section.}\label{nloptsolution}
\end{figure}

Figure~\ref{nloptsolution} shows an optics solution for the nonlinear energy collimation system. We use a mechanical spoiler and an absorber in between the two sextupoles. Two matching sections are included at the beginning and the end of the lattice. The skew sextupole parameters are shown in Table~\ref{skewsextparam}.

The collimation depth has been set to intercept beams with energy deviation larger than $1.3\%$ of the nominal energy. A horizontal spoiler and a horizontal absorber are used with half gap aperture $\approx 1$~mm.      

\section{Optical optimisation}

\begin{wrapfigure}{r}{0.5\columnwidth}
\centerline{\includegraphics[width=0.45\columnwidth]{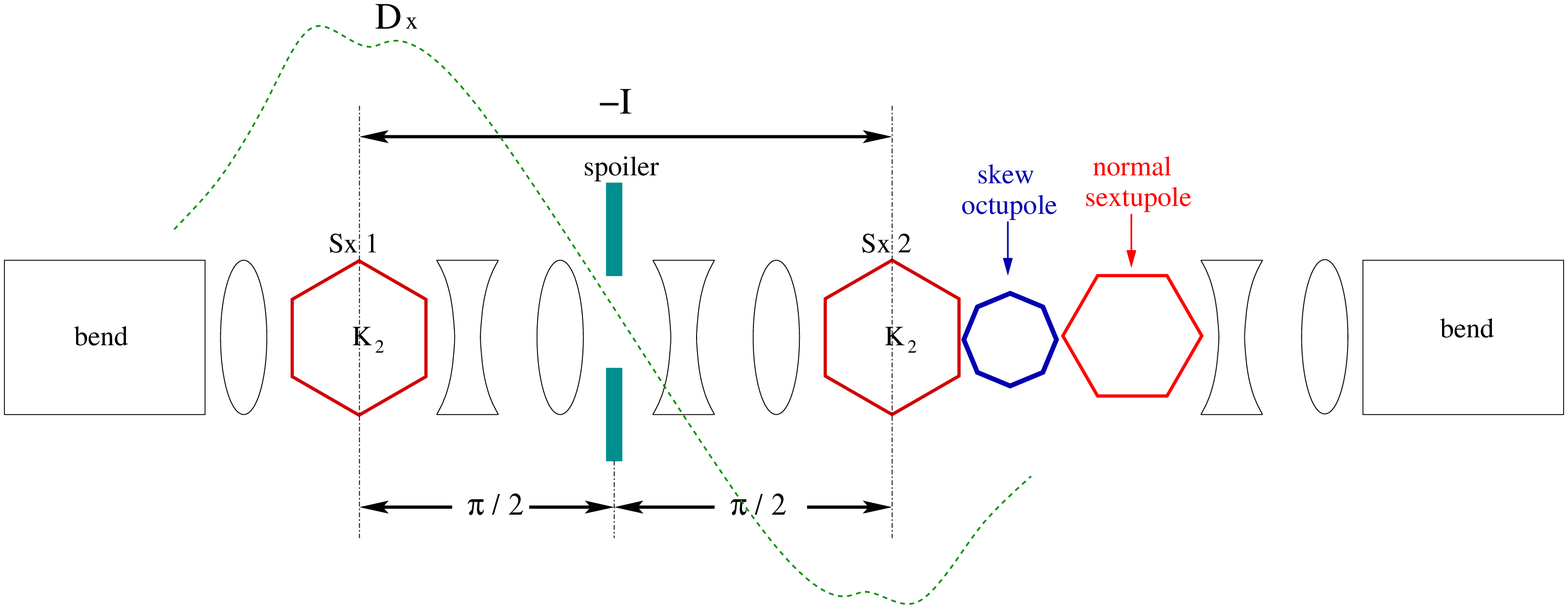}}
\caption{Schematic of a nonlinear collimation system using a pair of skew sextupoles of the same family, and adding a skew octupole and a normal sextupole for local correction of high order optical aberrations.}\label{nlcollopt}
\end{wrapfigure}

\label{opticsoptimisation}
Remnant higher order optical aberrations, mainly second and third order chromatic and chromo-geometric aberrations, are not effectively cancelled by the previous skew sextupole pair scheme ($s1$-$R(s1\rightarrow s2)$-$s2$, $R(s1\rightarrow s2)=-I$). This limits the luminosity performance. For the optimisation of the system, in order to cancel higher order aberrations we have added a skew octupole and a normal sextupole downstream of the second skew sextupole. Figure~\ref{nlcollopt} shows the schematic of the optimised lattice configuration. The strengths of the two additional nonlinear magnets have been calculated using the optimisation code MAPCLASS \cite{Rogelio}. 

In order to look for the optimum nonlinear magnet strengths we have to take into account the balance between the increase of the beam spot size for spoiler survival in case of beam impact, and an acceptable luminosity performance during normal beam operation. As condition, the luminosity loss $\Delta \mathcal{L}/\mathcal{L}_0 \lesssim 2\%$, with $\mathcal{L}_0 \simeq 6 \times 10^{34}$~cm$^{-2}$s$^{-1}$ the nominal CLIC peak luminosity.   

\section{Beamline performance}

Multiparticle tracking simulations have been performed to study the beam transport in the CLIC BDS with the nonlinear energy collimation system. For this study 50000 macroparticles were tracked through the BDS, simulating a beam with zero mean energy offset and with a uniform energy distribution (centred at the nominal beam energy 1500~GeV) with $1\%$ full energy spread. For the transverse phase space Gaussian beam distributions have been assumed. The code MAD \cite{MAD} has been used for this tracking study.  

\begin{figure}
\centerline{\includegraphics[width=0.50\columnwidth]{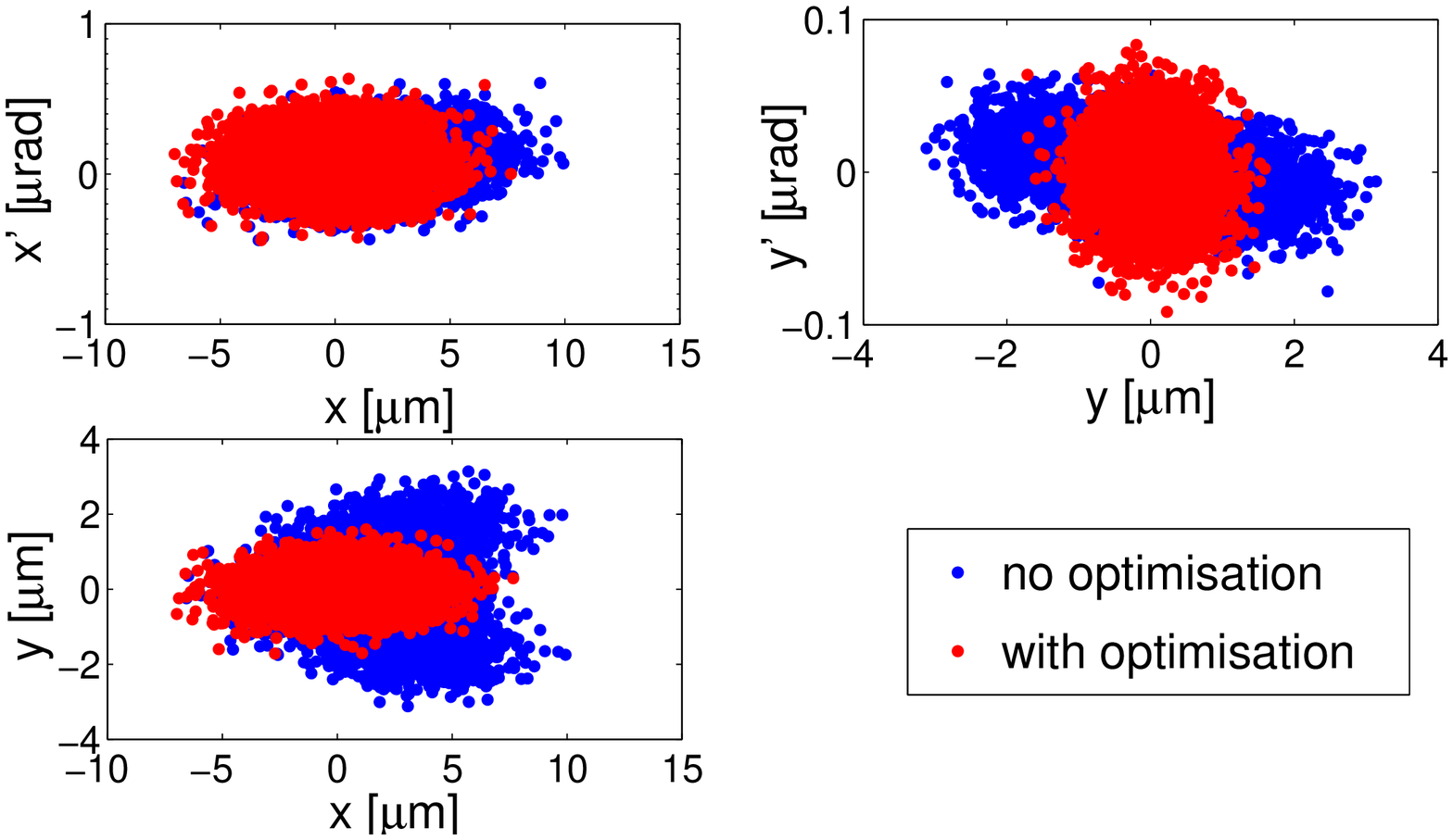} 
\includegraphics[width=0.50\columnwidth]{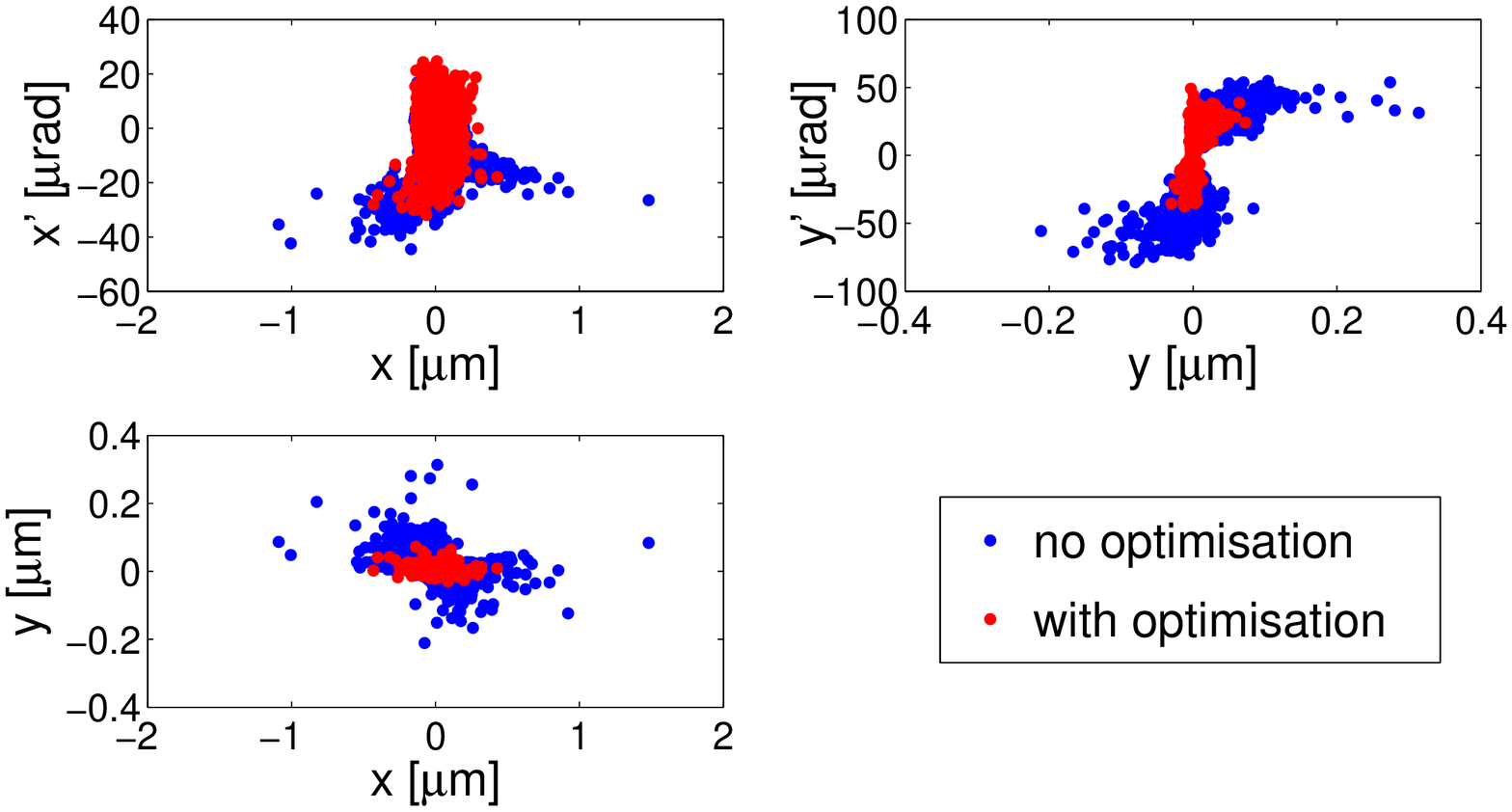}}
\caption{Transverse phase space of a CLIC beam at the exit of the nonlinear collimation section (Left) and at the IP (Right), with and without optical optimisation.}\label{phasespace}
\end{figure}

Figure~\ref{phasespace} shows the transverse phase space at the exit of the nonlinear energy collimation system and at the IP for the cases with and without optimisation. After optimisation (see Section~\ref{opticsoptimisation}) the beam tails are reduced and the beam core is more compact and much less distorted. 

\subsection{Luminosity}


\begin{wrapfigure}{r}{0.5\columnwidth}
\centerline{\includegraphics[width=0.5\columnwidth]{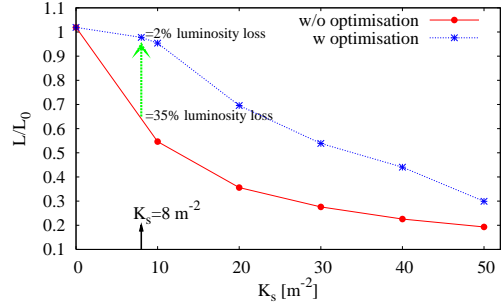}}
\caption{Relative peak luminosity as a function of the skew sextupole strength for the cases with and without optical optimisation.}\label{lumicompara}
\end{wrapfigure}

The luminosity has been computed at the IP using the beam-beam interaction code GUINEA-PIG \cite{GUINEA}. It is necessary to point out that to evaluate the luminosity performance of the lattice we have removed the aperture limitation in tracking to allow all particles reach the IP. Figure~\ref{lumicompara} shows the relative peak luminosity as a function of the integrated strength of the skew sextupoles. The cases with and without optimisation are compared. The luminosity is degraded quickly as the skew sextupole strength increases. The cancellation of high order aberrations using two additional nonlinear magnets (a skew octupole and a normal sextupole) helps to significantly improve the luminosity. For $K_s=8$~m$^{-2}$, without nonlinear optimisation, we obtain $\Delta \mathcal{L}/\mathcal{L}_0 \approx 35\%$. For $K_s=8$~m$^{-2}$, in the case of nonlinear optimisation with $K$(skew octupole)$=-2400$~m$^{-3}$ and $K$(normal sextupole)$=-0.4$~m$^{-2}$, we obtain $\Delta \mathcal{L}/\mathcal{L}_0 \approx 2\%$, i.e. $\mathcal{L}=5.8 \times 10^{34}$~cm$^{-2}$s$^{-1}$.    

\subsection{Emittance growth due to synchrotron radiation}
\label{SReffects}

The emittance growth due to synchrotron radiation (SR) emission must be constrained within tolerable levels. 

For a given lattice the horizontal emittance growth due to incoherent SR can be evaluated using the following expression \cite{Sands}:

\begin{equation}
\Delta (\gamma \epsilon_x) \simeq (4.13 \times 10^{-8}~{\textrm m}^2 \textrm{GeV}^{-6}) E^6 I_5 \,\,,
\label{emittancegrowth}
\end{equation}      

\noindent as a function of the beam energy $E$ and the so-called radiation integral $I_5$, which is defined as \cite{Helm},

\begin{equation}
I_5 = \int_{0}^{L} \frac{\cal H}{\left| \rho^{3}_{x} \right|} \,  ds =\sum_{i} L_i \frac{\langle {\cal H} \rangle_i}{\left| \rho^{3}_{x,i} \right|} \; ,
\label{nl:27}
\end{equation}

\noindent where the sum runs over all bending magnets, with bending radius $\rho_i$, length $L_i$, and the average of the function ${\cal H}$, which is defined by:

\begin{equation}
{\cal H}=\frac{D^2_x + (D'_x \beta_x + D_x \alpha_x)^2}{\beta_x}\,\,.
\end{equation} 


The beam core luminosity loss can be estimated from: 

\begin{equation}
\Delta \mathcal{L}/\mathcal{L}_0 = 1-1/\sqrt{ 1 +\Delta (\gamma \epsilon_x)/(\gamma \epsilon_{x})}\,\,. 
\label{eqlumiloss}
\end{equation}

For the design of the nonlinear collimation lattice we consider the following condition for the luminosity loss due to SR effects: $\Delta \mathcal{L}/\mathcal{L}_0 \lesssim 2\%$. This translates into the limit $I_5 \lesssim 6\times 10^{-20}$~m$^{-1}$. 

\noindent Table~\ref{emittgrowth} shows the values of $I_5$, $\Delta (\gamma \epsilon_x)/(\gamma \epsilon_{x})$ and $\Delta \mathcal{L}/\mathcal{L}_0$ calculated considering only the contribution from the collimation system and the contribution from the total BDS (collimation and final focus). The results for the nonlinear collimation system are compared with those for the baseline linear case. For the nonlinear collimation system the emittance growth is approximately four times lower than for the baseline linear system.   
     
\begin{table}
\centerline{\begin{tabular}{|l|c|c||c|c|}
\hline
{} & \multicolumn{2}{c||}{\bf CLIC BASELINE COLL.} & \multicolumn{2}{c|}{\bf CLIC NONLINEAR COLL.} \\
\hline 
Variable & (I) Coll. system & (II) Total BDS & (I) Coll. system & (II) Total BDS \\
\hline \hline
$I_5$ [m$^{-1}$] & $1.9 \times 10^{-19}$ & $3.8 \times 10^{-19}$ & $4.7 \times 10^{-20}$ & $2.4 \times 10^{-19}$ \\
$\Delta \epsilon_x/\epsilon_x$ [$\%$] & 13.5 & 27.3 & 3.3 & 17.1 \\
$\Delta \mathcal{L}/\mathcal{L}$ [$\%$] & 6.1 & 11.4 & 1.6 & 7.6 \\
\hline
\end{tabular}}
\caption{Values of the radiation integral $I_5$, emittance growth and luminosity loss due to incoherent SR effects taking into account the contribution from (I) only the collimation system and (II) the total BDS.}
\label{emittgrowth}
\end{table}


\subsection{Beam size at spoiler position}

From tracking simulations we have evaluated the transverse spot size, $\sqrt{\sigma_x \sigma_y}$, and the corresponding beam peak density (per bunch) at the spoiler position, $\rho=N/(2\pi\sigma_x \sigma_y)$, with $N$ the number of particles per bunch. 

\begin{figure}
\centerline{\includegraphics[width=0.5\columnwidth]{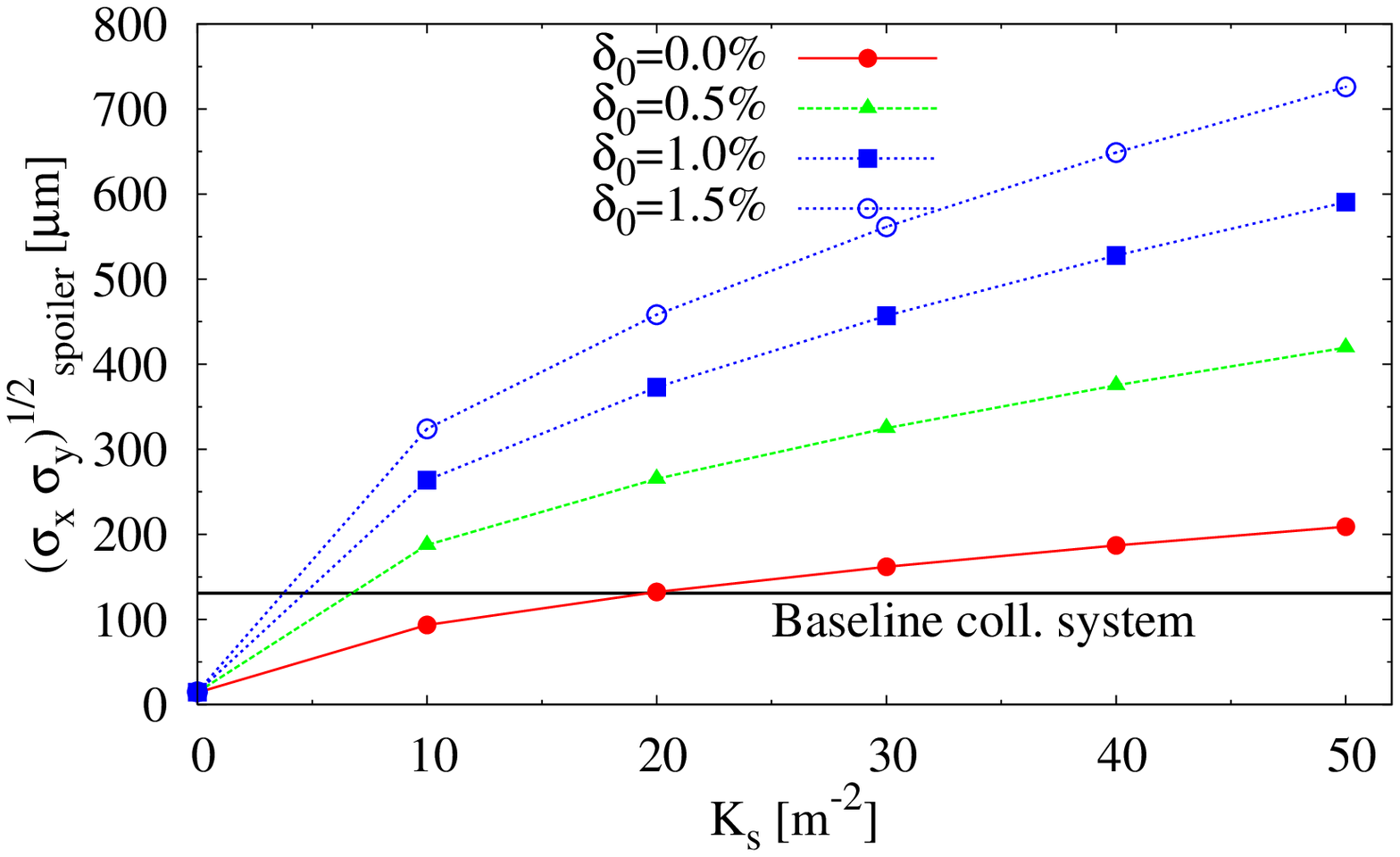}
\includegraphics[width=0.5\columnwidth]{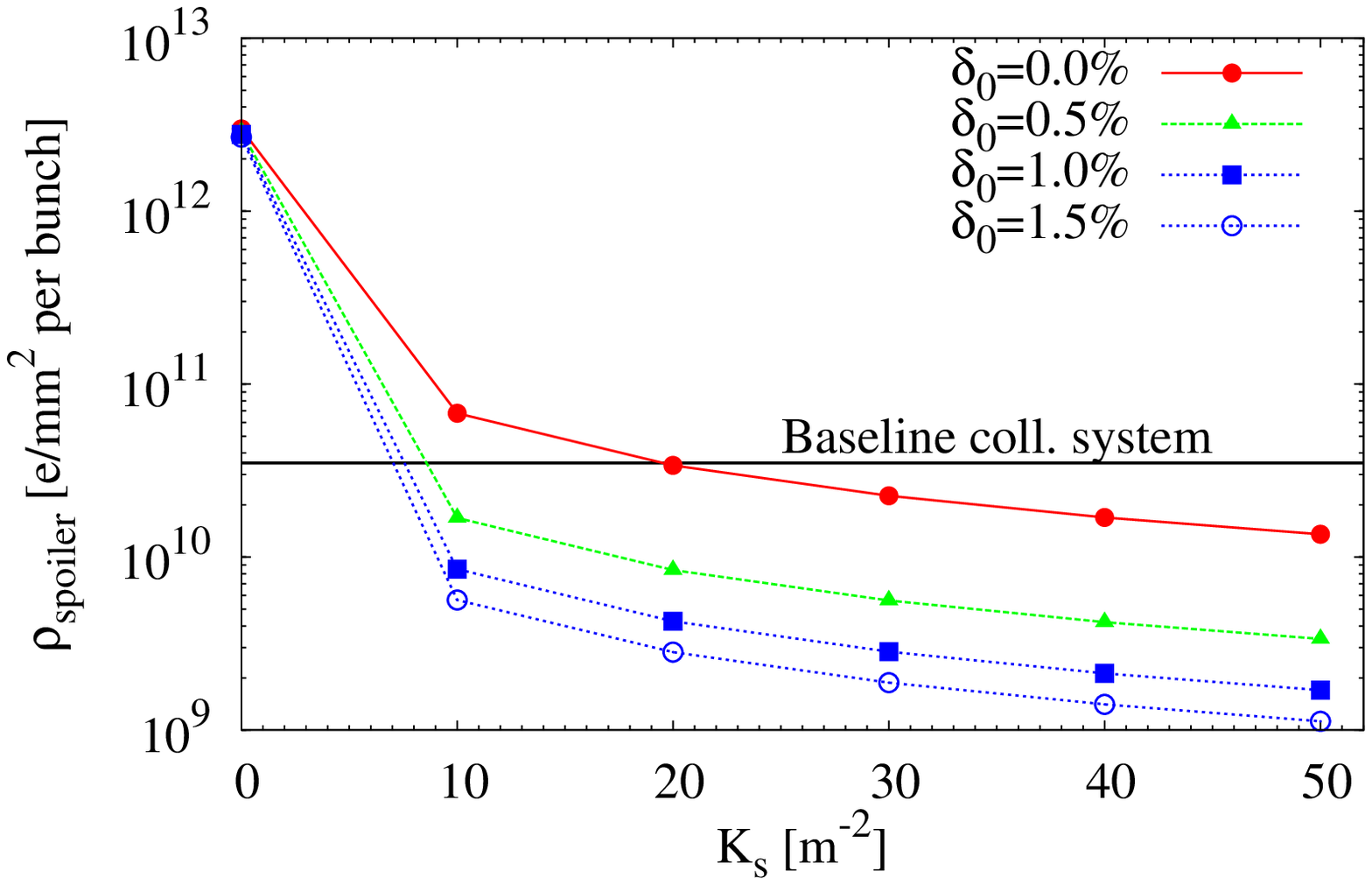}}
\caption{Transverse beam spot size (Left) and transverse beam peak density (Right) at the spoiler position versus the integrated skew sextupole strength, for different mean beam energy offsets from the nominal energy: $\delta_0\equiv \Delta E/E_0=0\%, 0.5\%, 1.0\%, 1.5\%$.}\label{density}
\end{figure}

Figure~\ref{density} shows the transverse spot size and the transverse beam peak density at spoiler position as a function of the skew sextupole strength for different mean energy offsets. The results are compared with the values for the case of the baseline linear collimation system (black solid line): $\sqrt{\sigma_x \sigma_y}=130.8$~$\mu$m and $\rho=3.5\times 10^{10}$ electrons (positrons) mm$^{-2}$ per bunch. For instance, in the case of $1.5\%$ mean energy offset, the nonlinear collimation system increases 2 times the beam spot size (reduces 4 times the transverse beam peak density) at the energy spoiler with respect to the baseline linear collimation system.

\section{Conclusions}
The increase of the transverse beam size at the collimators using nonlinear elements is a potential solution to guarantee the survival of the CLIC energy collimators in case of impact by a full bunch train. 

For CLIC we have presented the design of an alternative nonlinear energy collimation system based on a skew sextupole pair. Conditions for effective cancellation of optical aberrations of the lattice have been discussed. After beam optics optimisation, beam tracking simulations have shown an acceptable luminosity performance. Simulations have also shown a significant decrease of the transverse beam peak density at the spoiler position for beam energy offset $>1\%$, thus reducing the risk of material damage to the mechanical spoiler and absorber due to direct beam impact. 

In this paper we have presented a nonlinear energy collimation system for CLIC. However, this system is based in a general nonlinear optical scheme and could be adapted to other high energy colliders.  

Further studies include the investigation of a more compact optics design.    

\section{Acknowledgements}
This study is supported by grant FPA2010-21456-C02-01 from Ministerio de Ciencia e Innovaci\'on, Spain.


\begin{footnotesize}


\end{footnotesize}


\end{document}